\documentclass{ws-procs975x65}
\def\beq{\begin{equation}}
\def\eeq{\end{equation}}
\begin{document}

\title{Effective potential of particles in the oblique black hole magnetosphere}
\author{Ond\v{r}ej Kop\'{a}\v{c}ek and Vladim\'{i}r Karas}

\address{Astronomical Institute, Academy of Sciences,\\
Bo\v{c}n\'{i}~II~1401/1a, CZ-141\,31~Prague, Czech~Republic\\
E-mail: kopacek@ig.cas.cz\\
http://astro.cas.cz}

\begin{abstract}
Dynamics of charged matter in the oblique black hole magnetosphere is investigated. In particular, we adopt a model consisting of a rotating black hole embedded in the external large-scale magnetic field that is inclined arbitrarily with respect to the rotation axis. Breaking the axial symmetry appears to have profound consequences regarding the dynamics of particles and it also poses some methodological difficulties. In this contribution we discuss the applicability of the method of effective potential for the non-axisymmetric model and show that it may only be applied in the appropriate reference frame.
\end{abstract}

\keywords{black holes, magnetosphere, non-axisymmetry, dynamics of ionised matter, effective potential}

\bodymatter

%%%%%%%%%%%%%%%%% now a standard article style for the most part

\section{Introduction}
We investigate the dynamics of charged particles exposed to the strong gravitational and electromagnetic fields in the particular model of black hole magnetosphere. Previously, we studied an axisymmetric version of this model and successfully employed the method of the effective potential for the analysis.\cite{kovar08,kovar10,kopacek10} Nevertheless, if the axial symmetry breaks, the application of this tool and construction of the effective potential become problematic.\cite{kopacek14,kopacek15} In this paper we further discuss the applicability of the effective potential in the general non-axisymmetric case. More detailed introduction and astrophysical motivation for the model is given in Refs.~(\refcite{kopacek10,kopacek14}). 

\section{Inclined Magnetosphere of Rotating Black Hole}
Kerr metric  describing the geometry of the spacetime around a rotating black hole of mass $M$ and spin $a$ may be expressed in Boyer-Lindquist coordinates $x^{\mu}= (t,\:r, \:\theta,\:\varphi)$ as follows:\cite{mtw}
\begin{equation}
\label{metric}
ds^2=-\frac{\Delta}{\Sigma}\:[dt-a\sin{\theta}\,d\varphi]^2+\frac{\sin^2{\theta}}{\Sigma}\:[(r^2+a^2)d\varphi-a\,dt]^2+\frac{\Sigma}{\Delta}\;dr^2+\Sigma d\theta^2,
\end{equation}
where $\Delta(r)\equiv r^2-2Mr+a^2$ and $\Sigma(r,\theta)\equiv r^2+a^2\cos^2\theta$.

Geometrized units are used throughout the paper. Values of basic constants thus equal unity, $G=c=k=k_C=1$.

A test-field solution to Maxwell equations corresponding to the aligned magnetic field (of the asymptotic strength $B_z$) on the Kerr background  was derived by Wald.\cite{wald74} This solution was later generalized by Bi\v{c}\'{a}k and Jani\v{s}\cite{bicak85} to describe the field which is arbitrarily  inclined with respect to the rotation axis (direction of the magnetic field is then specified by two independent components, $B_z$ and $B_x$). Resulting vector potential of the electromagnetic field $A_{\mu}=(A_t,A_r,A_{\theta},A_{\varphi})$ is given explicitly by Eq.~(A4) of Ref.~(\refcite{bicak85}).   

% We note that in our previous studies \cite{karas12,karas09} we have investigated the structure of the electromagnetic field arising in this setup enriched by an extra ingredient: uniform motion of the black hole in arbitrary direction with respect to (arbitrarily inclined) magnetic field. However, here we consider the inclination of the magnetic field but not the boost of the black hole.

\section{Effective Potential in General Relativity}
% Considering the dynamics of charged test particle exposed to the external gravitational and electromagnetic fields in the classical framework, the method of effective potential is often applied. Analysis of the effective potential $V_{\rm eff}$ provides valuable overall information about the dynamics without the need of actual integration of particular trajectories. Its significance arises from the defining relation $\epsilon-V_{\rm eff}=T$ where $\epsilon$ is the classical total energy of the particle and $T$ represents its kinetic part. Thus the effective potential expresses the energy of the particle at which the turning points occur. It defines the boundary (in the extended configuration space) of the regions of allowed motion. In particular, it allows to locate stable circular orbits which correspond to the (local) minima of the effective potential. 

In general relativity we have no clear distinction between the kinetic and potential energy. Nevertheless, in many cases we may still derive function analogous to the classical effective potential. Such a function locates the turning points of motion and represents a boundary (in the extended configuration space) of allowed regions. Analysis of the effective potential $V_{\rm eff}$ provides valuable overall information about the dynamics without need of actual integration of particular trajectories and, most importantly, it allows to locate regions of stable orbits. Searching for the effective potential in the general case of a charged particle of the rest mass $m$ and charge $q$ in the spacetime with metric $g^{\mu\nu}$ and electromagnetic field $A_{\mu}$ we start from the Hamiltonian $\mathcal{H}$ expressed in canonical variables ($x^{\mu}$, $\pi_{\mu}$) whose conserved value is given by the normalization of the kinematic four-momentum $p^{\mu}$:\cite{mtw}
\begin{equation}
\label{fourmomentum} 2\mathcal{H}=g^{\mu\nu}p_{\mu}p_{\nu}=g^{\mu\nu}(\pi_{\mu}-qA_{\mu})(\pi_{\nu}-qA_{\nu})=-m^2.
\end{equation}

\subsection{Axisymmetric Magnetosphere}
In the special case of stationary and axisymmetric Kerr spacetime with additional electromagnetic test-field obeying the same symmetries (in which case $\pi_{\varphi}=L$ and $\pi_t=-E$ are constants of motion and system therefore has two degrees of freedom) we obtain from Eq.~(\ref{fourmomentum}) by straightforward manipulations 
\begin{equation}
\label{eff}
\Sigma\left(\frac{(p^{r})^2}{\Delta}+(p^{\theta})^2\right)=\alpha{}E^2+\beta{}E+\gamma,
\end{equation}
where
\begin{eqnarray}
\label{coeff}
\alpha&=&-g^{tt},\;\;\;\;\;\;\;\beta=2\left[g^{t\varphi{}}(L-qA_{\varphi})-g^{tt}qA_{t}\right],\\
\gamma&=&-g^{\varphi\varphi}(L-qA_{\varphi})^2-g^{tt}q^2A_{t}^2+2g^{t\varphi{}}qA_{t}(L-qA_{\varphi})-m^2.
\end{eqnarray}
Since both coefficients $\Sigma$ and $\Delta$ are positive above the outer horizon $r=r_+$ to which region we restrict our study, the zero point of the left-hand side of Eq.~(\ref{eff}) occurs at the simultaneous turning point of motion in both the radial and latitudinal directions and it defines the boundary of allowed motion. Function which specifies the value of energy corresponding to the turning point can be regarded as a generalization of the classical effective potential $V_{\rm eff}$. We can therefore express the two-dimensional effective potential as $V_{\rm eff}(r,\theta)=\left(-\beta+\sqrt{\beta^2-4\alpha\gamma}\right)/2\alpha$, where the positive root of quadratic equation has to be chosen to correspond with the future-pointing four-momentum.\cite{mtw} Since $\alpha>0$ above the horizon the motion is allowed just if $E\geq V_{\rm eff}$.

Method of effective potential has been applied to locate confinements (both equatorial and off-equatorial) of charged matter in several stationary and axisymmetric models in our previous works.\cite{kopacek10, kovar10, kovar08} Potential $V_{\rm eff}$ was investigated as a function of two configuration variables $r$ and $\theta$, angular momentum $L$ of the particle and parameters of the given system. 

\subsection{Oblique Magnetosphere}
The question arises whether we could also apply the method of effective potential for the stationary system of three degrees of freedom in which the axial symmetry is broken and $A_{\mu}=A_{\mu}(r,\theta,\varphi)$ but $g_{\mu\nu}=g_{\mu\nu}(r,\theta)$. In this case the trajectory manifold spans five dimensions out of total eight dimensions of the phase space. The effective potential reduces the number of dimensions by imposing the constraint of type $(p^{\mu})^2=0$ which locates the turning point in given direction. Here we seek the simultaneous turning point in all three directions $r$, $\theta$ and $\varphi$ which would result in a two-dimensional submanifold. For a fixed value of $\varphi$ we should therefore obtain one-dimensional isopotential curves specifying the allowed region in a given meridional plane described by coordinates $r, \theta$ as we previously did in the case of axisymmetric systems. Indeed, we can derive the expression formally analogous to Eq.~(\ref{eff}):

\begin{equation}
\label{eff2}
\Sigma\left(\frac{(p^{r})^2}{\Delta}+(p^{\theta})^2\right)+g_{\varphi\varphi}(p^{\varphi})^2=\alpha^{\star}E^2+\beta^{\star}{}E+\gamma^{\star},
\end{equation}
where the coefficients are now given as
\begin{eqnarray}
\label{coeff2}
\alpha^{\star}&=&-g^{tt}\left(1+g^{t\varphi}g_{t\varphi}\right),\;\;\;\;
\beta^{\star}=2\left[g_{t\varphi}(g^{t\varphi})^2(\pi_{\varphi}-qA_{\varphi})-g^{tt}qA_{t}(1+g^{t\varphi}g_{t\varphi})\right]\\
{\nonumber}\gamma^{\star}&=&-g^{\varphi\varphi}g^{t\varphi}g_{t\varphi}(\pi_{\varphi}-qA_{\varphi})^2-g^{tt}q^2A_{t}^2(1+g^{t\varphi}g_{t\varphi})+2(g^{t\varphi{}})^2g_{t\varphi}qA_{t}(\pi_{\varphi}-qA_{\varphi})-m^2.
\end{eqnarray}

\begin{figure}[ht]
\includegraphics[scale=0.41,trim=0mm 0mm 0mm 0mm,clip]{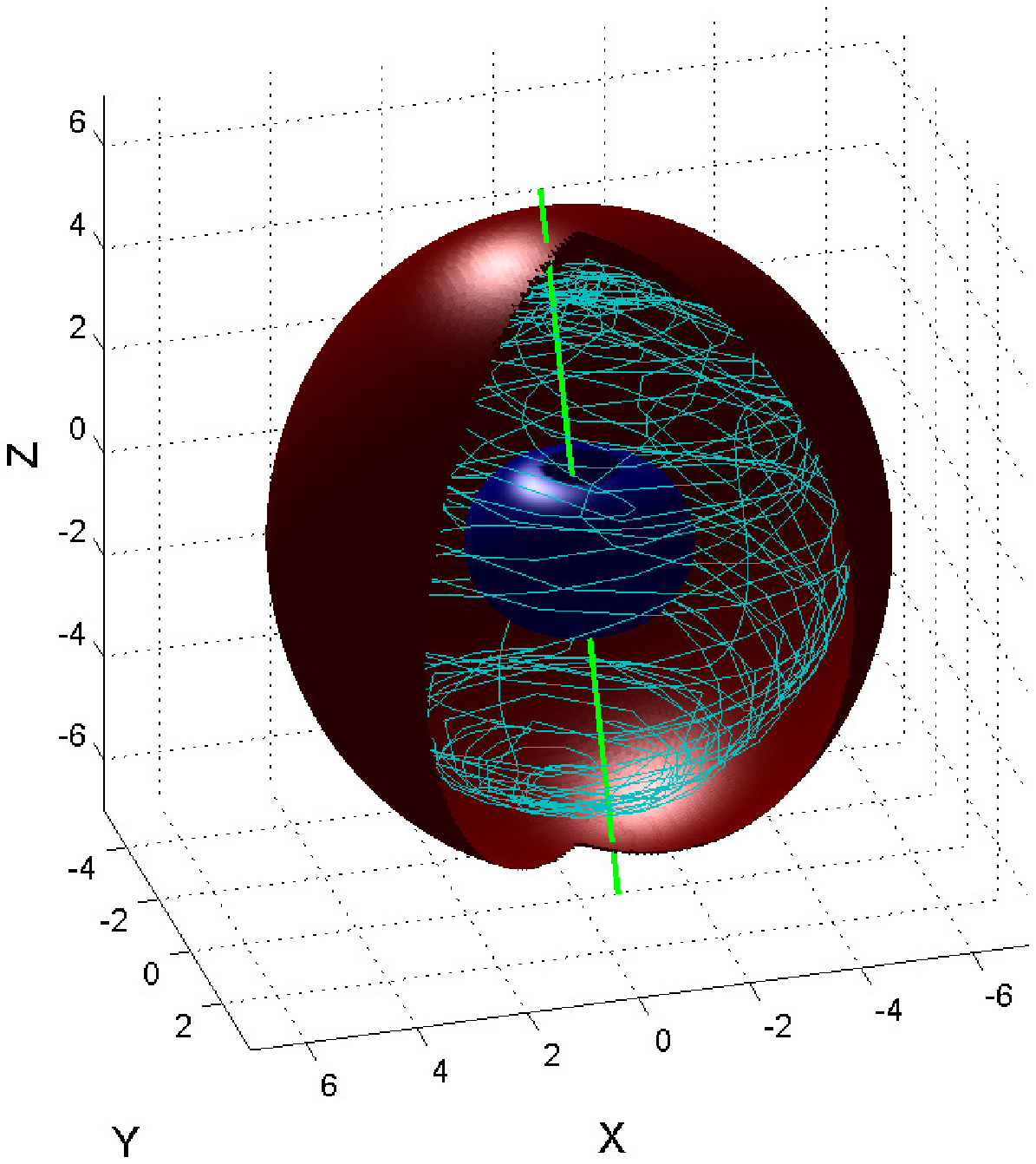}~~~~~
\includegraphics[scale=0.36,trim=0mm 0mm 0mm 0mm,clip]{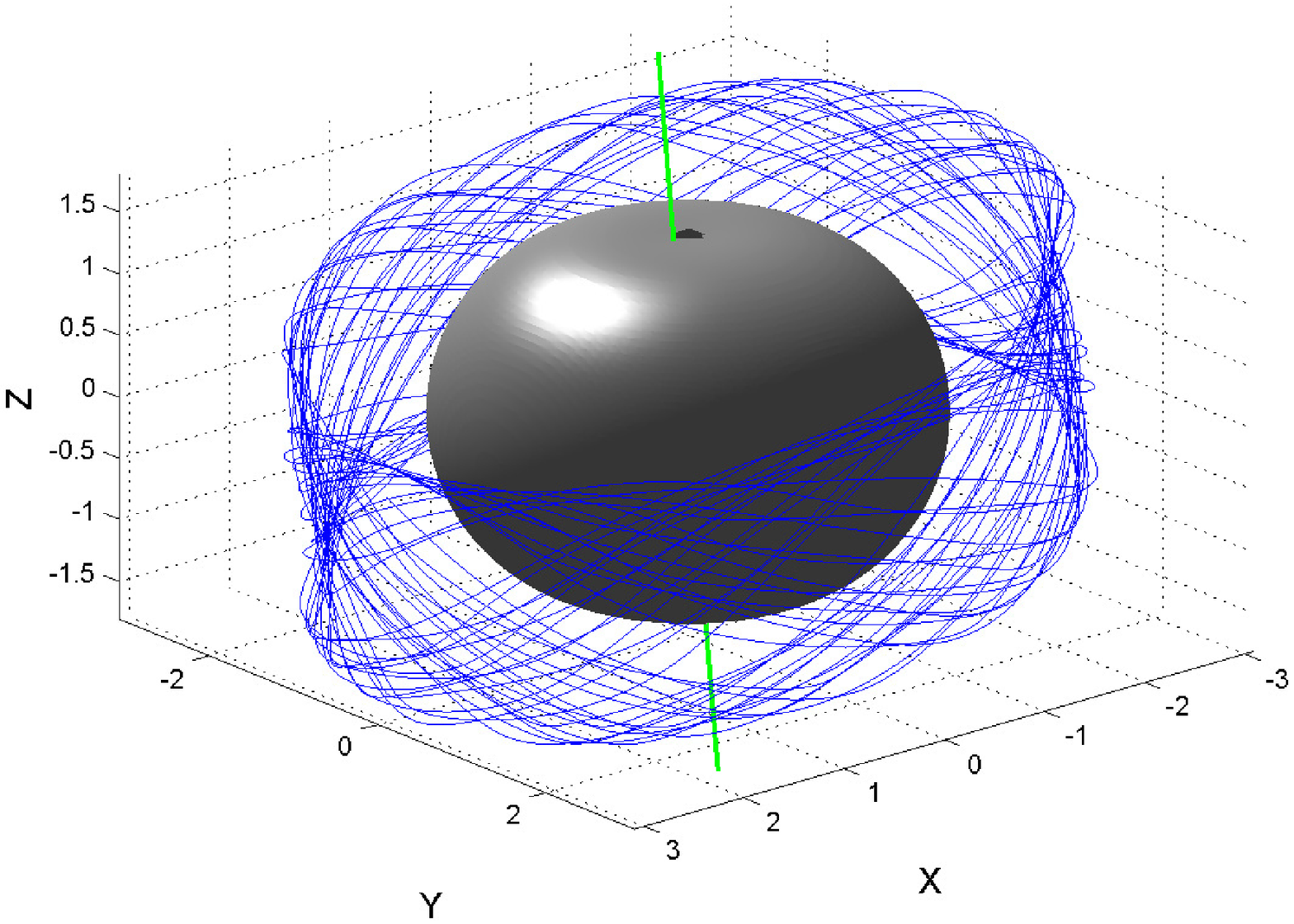}
\caption{Particular trajectory (with parameters $E=1.24$, $qB_z=1$, $B_x/B_z=0.1$, $a=0.9$, and the initial condition $\theta(0)=\pi/2$, $\varphi(0)=0$, $u^r(0)=0$ and $\pi_{\varphi}(0)=5$) is  observed in both the static frame (left panel) and Boyer-Lindquist coordinate frame (right panel). The former is bounded by the corresponding isosurface of effective potential (red) while the inner blue-colored surface represents the ergosphere of the black hole inside which the static frame becomes unphysical. In the right panel the ergosphere is marked by the gray surface instead. Asymptotic direction of the magnetic field is indicated by the green line. Distances are scaled by black hole mass $M$.}
\label{effpot1}
\end{figure}

Left-hand side of Eq.~(\ref{eff2}) has the proper form necessary for expressing the effective potential ($g_{\varphi\varphi}$ is positive). Nevertheless, the coefficients $\beta^{\star}$ and $\gamma^{\star}$ depend on the azimuthal component of canonical momentum $\pi_{\varphi}$ which used to be the integral of motion $L$ in the axisymmetric system, however, here it changes along the trajectory. Evolution of $\pi_\varphi$ is not known a priori, and one has to integrate the equations of motion of given particle to reveal it. Therefore it is not possible to express the effective potential from the equation  (\ref{eff2}) as a function of $r$, $\theta$, $\varphi$ and parameters of the metric and electromagnetic field. Simultaneous turning points and the boundaries of allowed regions are actually not captured by this formula.

Nevertheless, it has been recently demonstrated\cite{epp13,epp14} that in the classical analogue of the investigated system one may proceed by leaving the coordinate basis and employing the appropriate reference frame. Indeed, switching from the Cartesian coordinate grid to the co-rotating frame allowed to express the effective potential of charged particles in the field of a rigidly rotating inclined magnetic dipole.\cite{epp13,epp14}

In our context this would mean to choose an observer with the orthonormal tetrad vectors $e_{(\mu)}^{\nu}$ which allow to express the boundaries of allowed motion of particle with mass $m$, charge $q$ and kinematical four-momentum $p_{(\mu)}=e_{(\mu)}^{\nu}p_{\nu}$ as follows:
\begin{equation}
\label{effpo2}
p^2_{(r)}+p^2_{(\theta)}+p^2_{(\varphi)}=\tilde{\alpha}E^2+\tilde{\beta}E+\tilde{\gamma}\geq 0,
\end{equation}
making sure that coefficients $\tilde{\alpha}$, $\tilde{\beta}$ and $\tilde{\gamma}$ only depend on configuration variables $r$, $\theta$, $\varphi$ and parameters of the system ($a$, $q$, $B_x$ and $B_z$). This relation follows directly from the covariance of expression (\ref{fourmomentum}) and orthonormality of the tetrad (i.e., $g^{(\mu)(\nu)}=\eta^{(\mu)(\nu)}$). Our search for the proper tetrad is naturally restricted to the class of stationary frames\cite{semerak93} characterized by the four-velocity of the form $e^{\mu}_{(t)}=u^{\mu}=(u^t,0,0,u^{\varphi})$. Prominent examples of stationary frames in Kerr spacetime are those carried by zero angular momentum observer (also denoted as locally non-rotating frame\cite{bardeen72}) and geodesic frame of Keplerian observer orbiting the black hole on stable circular orbits in equatorial plane. 

In particular, the time component of the canonical four-momentum  measured in a general stationary frame is expressed as  $\pi_{(t)}=u^t\pi_t+u^{\varphi}\pi_{\varphi}=-u^tE+u^{\varphi}\pi_{\varphi}$. It appears, however, that no stationary frame can eliminate $\pi_{\varphi}$ appearing in this formula from the coefficients $\tilde{\beta}$ and $\tilde{\gamma}$, and we end up in the same situation as we did in the Boyer-Lindquist coordinate basis with coefficients $\beta^{\star}$ and $\gamma^{\star}$ given by the relation~(\ref{coeff2}). The only solution is to additionally demand $u^{\varphi}=0$, i.e., to switch to the static frame with tetrad vectors\cite{semerak93}
\begin{align}
\label{statictetrad1}
e_{(t)}^{\mu}&= \left[\frac{\Sigma^{1/2}}{\chi},0,0,0\right],\;\;\;
e_{(r)}^{\mu}=\left[0,\frac{\Delta^{1/2}}{\Sigma^{1/2}},0,0\right],\\
e_{(\theta)}^{\mu}&=\left[0,0,\frac{1}{\Sigma^{1/2}},0\right],\;\;\;
\label{statictetrad4}
e_{(\varphi)}^{\mu}=\frac{\chi}{\sin\theta\Delta^{1/2}\Sigma^{1/2}}\left[\frac{-2aMr\sin^2\theta}{\chi^2},0,0, 1\right],
\end{align}
where $\chi^2\equiv\Delta-a^2\sin^2\theta$.

\begin{figure}[ht]
\includegraphics[scale=0.43,trim=0mm 0mm 0mm 0mm,clip]{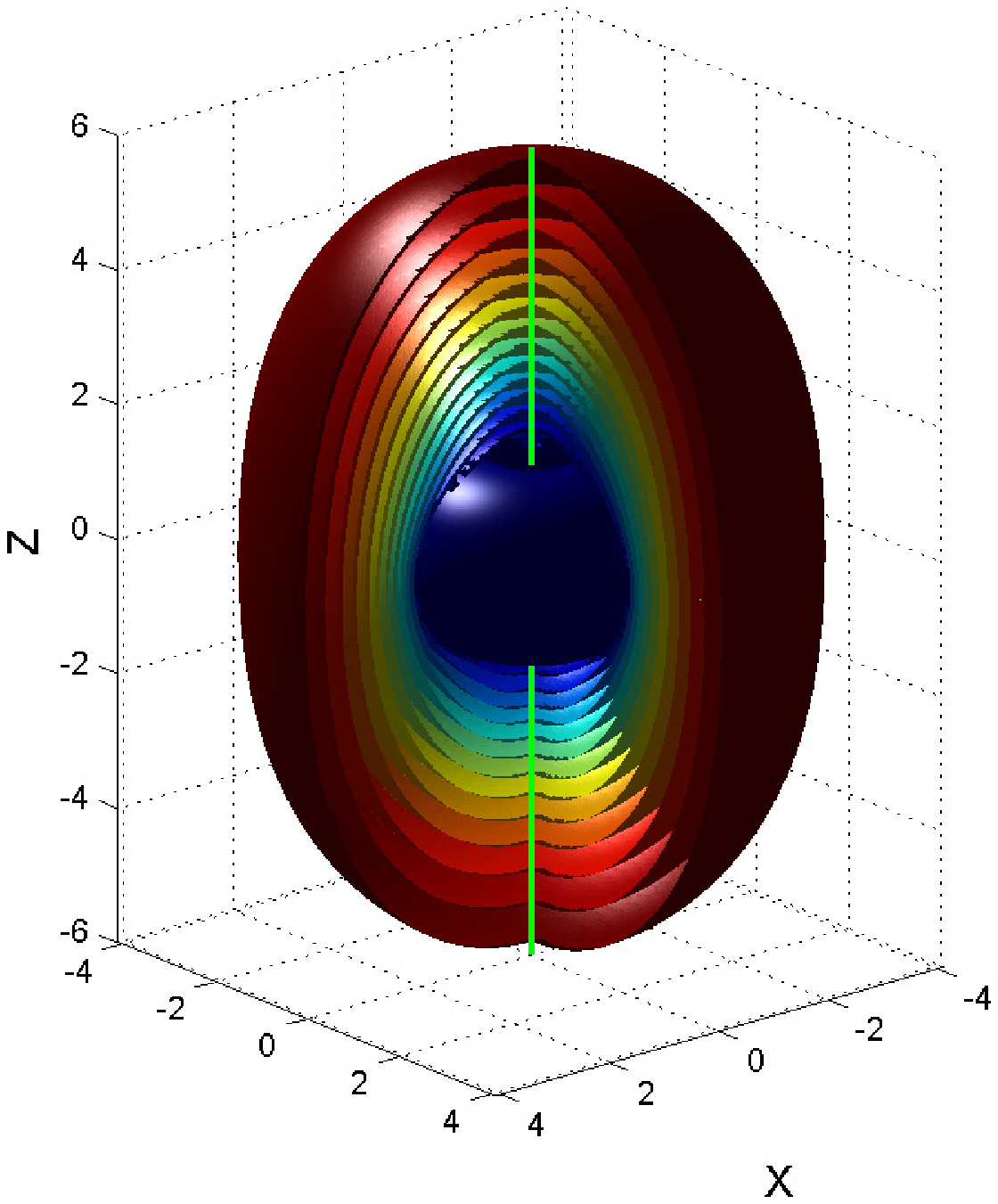}~~~
\includegraphics[scale=0.43,trim=0mm 0mm 0mm 0mm,clip]{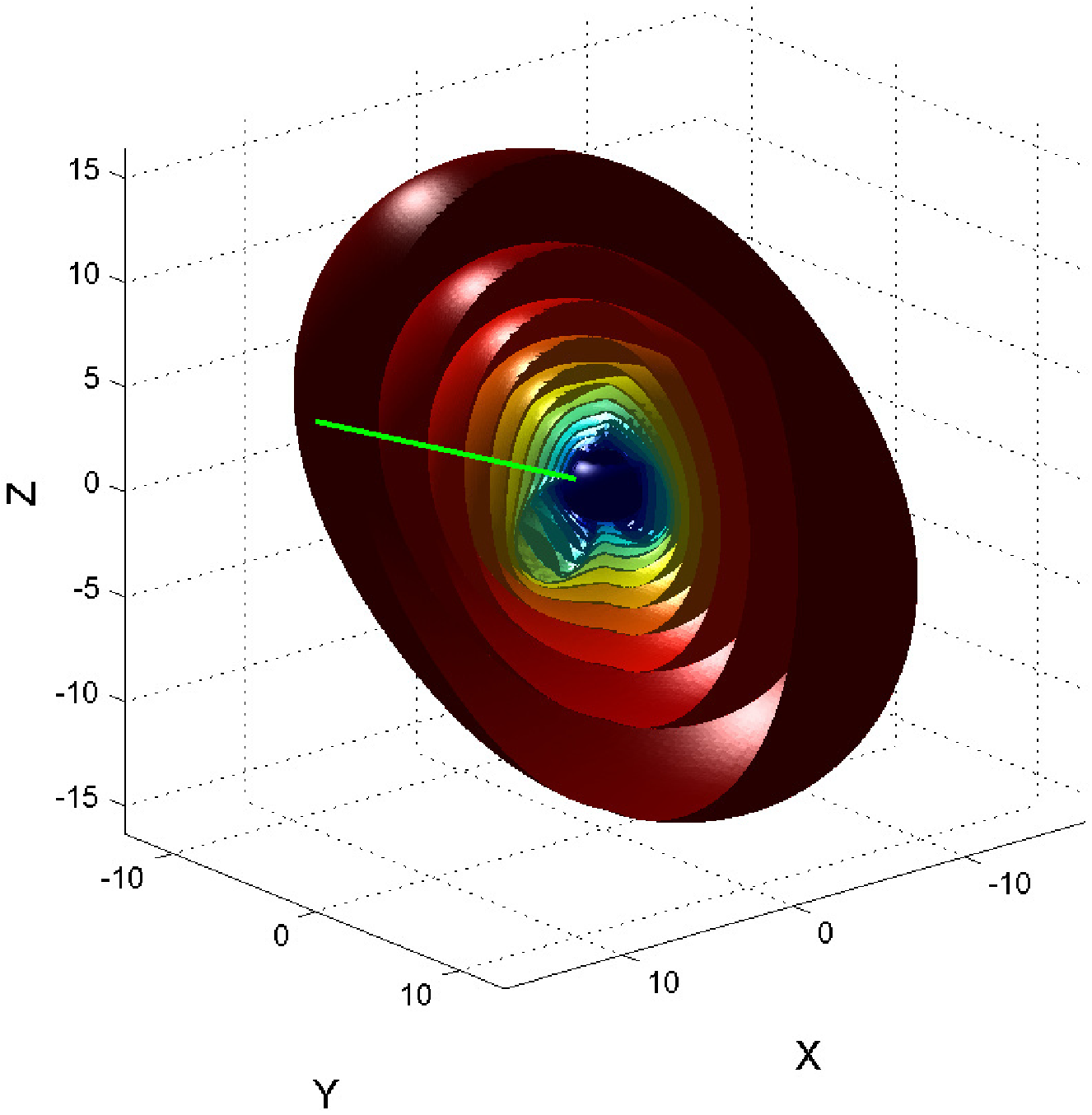}\\
\includegraphics[scale=0.43,trim=0mm 0mm 0mm 0mm,clip]{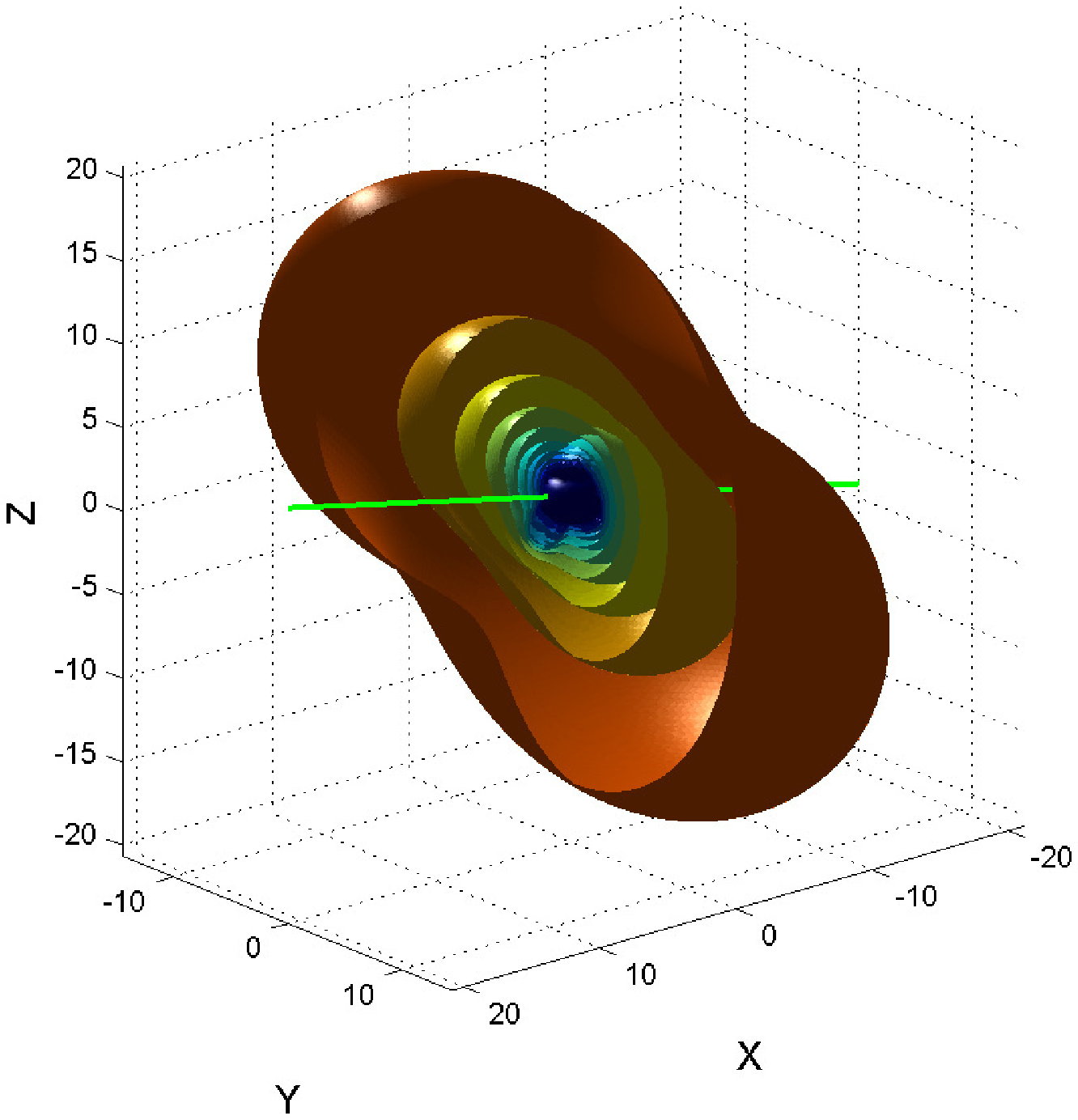}~~~
\includegraphics[scale=0.43,trim=0mm 0mm 0mm 0mm,clip]{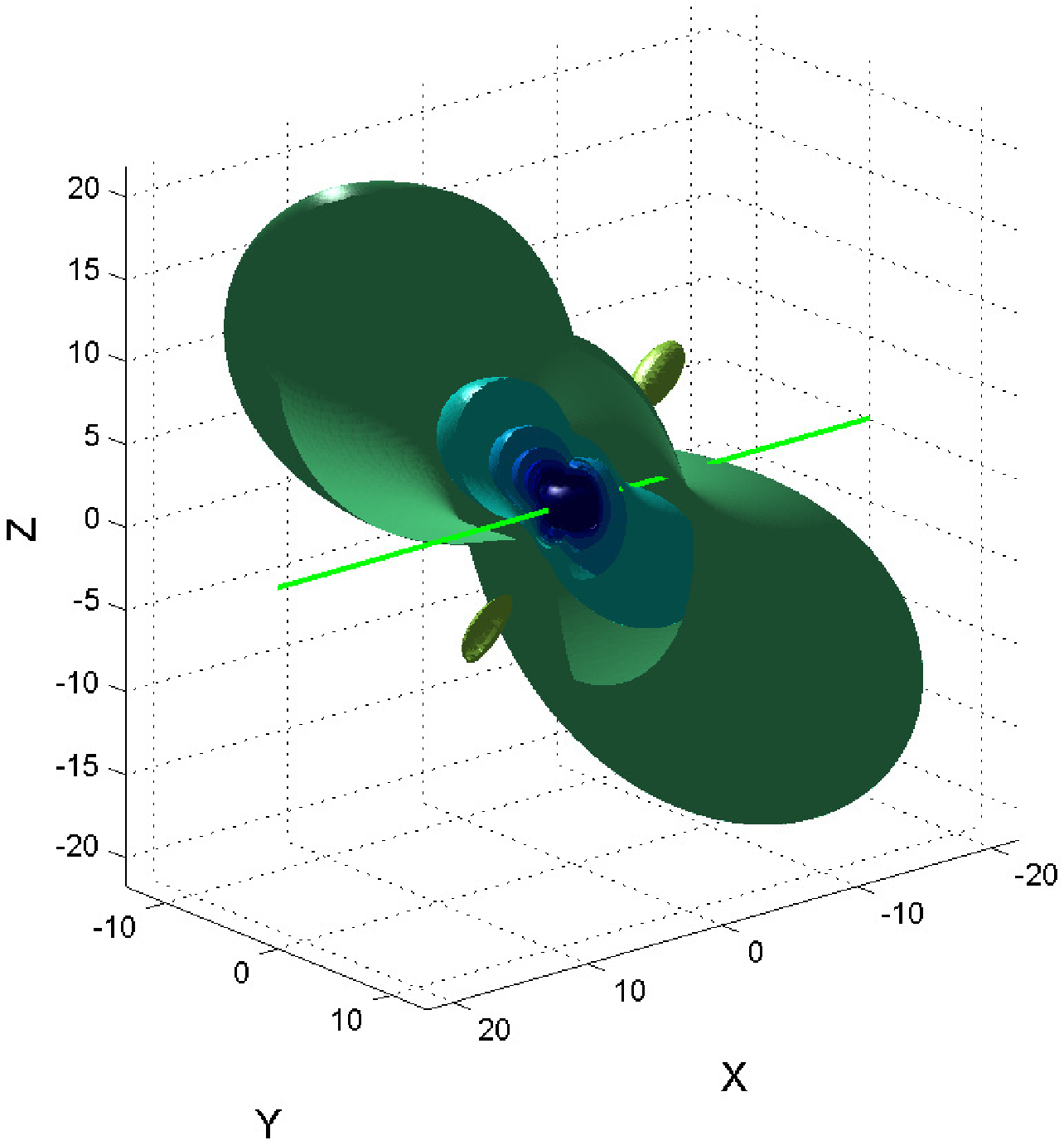}
\caption{Shape of the isosurfaces of the effective potential changes profoundly as the inclination of the magnetic field (green line) gradually increases. Common parameters of the system are $a=0.9$ and $qB=q\sqrt{B_x^2+B_z^2}=4$ while the inclination angle $\alpha\equiv\arctan(B_x/B_z)$ of the magnetic field gradually rises as $\alpha=0, \frac{3\pi}{8}, \frac{7\pi}{16}$ and $\frac{\pi}{2}$ (top left to bottom right in presented figures). Inner blue-colored surface corresponds to the ergosphere. Distances are scaled by the central mass $M$.}
\label{effpot2}
\end{figure}

This static frame allows to express the effective potential as $V_{\rm eff}(r,\theta,\varphi)=\left(-\tilde{\beta}+\sqrt{\tilde{\beta}^2-4\tilde{\alpha}\tilde{\gamma}}\right)/2\tilde{\alpha}$, where the coefficients are defined as
\begin{equation}
\tilde{\alpha}=\left[e^t_{(t)}\right]^2,\;\;\;\tilde{\beta}=2 q A_t e^t_{(t)},\;\;\;\tilde{\gamma}=q^2\left[e^t_{(t)}\right]^2A_t^2-m^2.
\end{equation}
Nevertheless, in the Kerr spacetime no observer may remain static inside the ergosphere whose boundary (corresponding to $\chi^2=0$) is defined by $r_{\rm{s}}=M+\sqrt{M^2-a^2\cos{\theta}^2}$. As a result the effective potential constructed in static frame is well-defined only outside the ergosphere. However, as demonstrated above, neither non-static frames with $u^{\varphi}\neq 0$, nor the Boyer-Lindquist coordinate basis itself allow to express the effective potential in the non-axisymmentric case of oblique magnetosphere of rotating black hole.

In Fig.~\ref{effpot1} we present a particular trajectory of charged particle in a slightly inclined magnetic field as viewed in the frame of a static observer and we show its boundary represented by the corresponding isosurface of the effective potential. This trajectory is also shown in the coordinate frame for comparison. In Fig.~\ref{effpot2} we observe how the shape of isosurfaces evolves as the inclinations of the magnetic field rises.

% \begin{align}
% \label{zamotetrad1}
%  e_{(t)}^{\mu}&= u^{\mu}=\frac{A^{1/2}}{\Delta^{1/2}\Sigma^{1/2}}\left[1,0,0,\Omega\right],\\
% e_{(r)}^{\mu}&=\left[0,\frac{\Delta^{1/2}}{\Sigma^{1/2}},0,0\right],\\
% e_{(\theta)}^{\mu}&=\left[0,0,\frac{1}{\Sigma^{1/2}},0\right],\\
% \label{zamotetrad4}
% e_{(\varphi)}^{\mu}&=\left[0,0,0, \frac{\Sigma^{1/2}}{A^{1/2}\sin\theta}\right],\\
% e^{(t)}_{\mu}&= u_{\mu}=\left[\frac{\Sigma^{1/2} \Delta^{1/2}}{A^{1/2}},0,0,0\right],\\
% e^{(r)}_{\mu}&=\left[0,\frac{\Sigma^{1/2}}{\Delta^{1/2}},0,0\right],\\
% e^{(\theta)}_{\mu}&=\left[0,0,\Sigma^{1/2},0\right],\\
% e^{(\varphi)}_{\mu}&=\frac{A^{1/2}\sin\theta}{\Sigma^{1/2}}\left[-\Omega,0,0,1\right],
% \end{align}
% 
% \begin{align}
% \label{statictetrad1}
% e_{(t)}^{\mu}&= u^{\mu}=\left[\frac{\Sigma^{1/2}}{\chi},0,0,0\right],\\
% e_{(r)}^{\mu}&=\left[0,\frac{\Delta^{1/2}}{\Sigma^{1/2}},0,0\right],\\
% e_{(\theta)}^{\mu}&=\left[0,0,\frac{1}{\Sigma^{1/2}},0\right],\\
% \label{statictetrad4}
% e_{(\varphi)}^{\mu}&=\frac{\chi}{\sin\theta\Delta^{1/2}\Sigma^{1/2}}\left[\frac{-2aMr\sin^2\theta}{\chi^2},0,0, 1\right],\\
% e^{(t)}_{\mu}&= u_{\mu}=\frac{\chi}{\Sigma^{1/2}}\left[1,0,0,\frac{2aMrsin^2\theta}{\chi^2}\right],\\
% e^{(r)}_{\mu}&=\left[0,\frac{\Sigma^{1/2}}{\Delta^{1/2}},0,0\right],\\
% e^{(\theta)}_{\mu}&=\left[0,0,\Sigma^{1/2},0\right],\\
% e^{(\varphi)}_{\mu}&=\left[0,0,0,\frac{\Sigma^{1/2}\Delta^{1/2}\sin\theta}{\chi}\right],
% \end{align}

\section{Conclusions}
We have constructed the effective potential for charged particles in the oblique black hole magnetosphere. In order to do so, the proper reference frame had to be employed. While in the classical analogue of investigated system the co-rotating frame was used in this context, we had to switch to the static frame instead. In this respect, the static frame in the Kerr geometry appears to represent the analogy of classical co-rotating frame. This is somewhat surprising as the the frame which is here usually associated with {\em co-rotation} is rather the locally non-rotating frame.\cite{bardeen72}

%In the following we plan to systematically search for the stationary points of the effective potential and discuss the stability of corresponding trajectories in order to locate stable confinements of ionised matter in the oblique magnetosphere of rotating black hole.

\section*{Acknowledgments}
Authors thank the Czech Science Foundation for support via the project GA\v{C}R 14-37086G and acknowledge the bilateral Czech-German cooperation project DAAD 15-14. Private communication with Vladimir Epp is highly appreciated.

\end{document}